\documentclass[10pt]{article}

\usepackage{amsmath}
\usepackage{amssymb}

\usepackage{graphicx}

\usepackage{cite}

\usepackage{color} 

\usepackage{hyperref}
\hypersetup{
    colorlinks=true,       
    linkcolor=magenta,          
    citecolor=red,        
    filecolor=magenta,      
    urlcolor=blue           
}

\usepackage[labelfont=bf,labelsep=period,justification=raggedright]{caption}

\usepackage{subcaption}
\makeatletter
\renewcommand{\@biblabel}[1]{\quad#1.}
\makeatother

\date{}

\begin{document}

\begin{flushleft}
{\Large
\textbf{A hierarchical network heuristic for solving the orientation problem in genome assembly}
}
\\
\bigskip

Karl R. B. Schmitt$^{124}$, 
Aleksey V Zimin$^{4\ast}$,
Guillaume Marca\c{c}s$^{4}$,
James A Yorke$^{234}$,
Michelle Girvan$^{34}$
\\
\bigskip
{1} Department of Mathematics and Computer Science, Valparaiso University, Valparaiso, Indiana, USA
\\
{2} Department of Mathematics, University of Maryland, College Park, Maryland, USA
\\
{3} Department of Physics, University of Maryland, College Park, Maryland, USA
\\
{4} Institute for Physical Science and Technology, University of Maryland, College Park, Maryland, USA
\\
$\ast$ Corresponding Author, E-mail: alekseyz@ipst.umd.edu
\end{flushleft}

\section*{Abstract}

In the past several years, the problem of genome assembly has received considerable attention from both biologists and computer scientists.  An important component of current assembly methods is the scaffolding process. This process involves building ordered and oriented linear collections of contigs (continuous overlapping sequence reads) called scaffolds and relies on the use of mate pair data.  A mate pair is a set of two reads that are sequenced from the ends of a single fragment of DNA, and therefore have opposite mutual orientations. When two reads of a mate-pair are placed into two different contigs, one can infer the mutual orientation of these contigs. While several orientation algorithms exist as part of assembly programs, all encounter challenges while solving the orientation problem due to errors from mis-assemblies in contigs or errors in read placements. In this paper we present an algorithm based on hierarchical clustering that independently solves the orientation problem and is robust to errors. We show that our algorithm can correctly solve the orientation problem for both faux (generated) assembly data and real assembly data for \textit{R. sphaeroides bacteria}. We demonstrate that our algorithm is stable to both changes in the initial orientations as well as noise in the data, making it advantageous compared to traditional approaches.

\section*{Author Summary}
Constructing an organism's entire DNA sequence from raw genome sequencing data, like the data produced in the Human Genome Project, is a challenging task. The type of data generated in the sequencing process has changed substantially over the years as a result of various technological improvements.  The computer programs that convert such data into assembled sequencing must continuously be revised to keep pace with the changing nature of the data.  This paper builds upon current methods from the emerging field of network science to develop a new way of analyzing and correcting sequencing data.  We show that our algorithm is both more robust to erroneous data, and more accurate overall, compared to current techniques.

\section{Introduction}
\label{sec:intro3}
In the late 1990's and early 2000's, the Human Genome Project made headlines worldwide. The goal of the project was to determine the sequence of chemical base pairs in human DNA \cite{huangmurray}. In that era it took years to completely sequence an individual organism's DNA structure \cite{lander2001,lander2011}. Now, with a new generation of genome sequencers that recombine different types of data using revised algorithms \cite{pop2002}, it is possible to sequence a genome in several weeks or less\cite{schatz2010}.  However, despite these improvements, current technologies for genome sequencing still involve significant errors.  In this paper we offer an improvement on current sequencing algorithms.

At the most basic level sequencing a genome is a step by step method for solving a puzzle. Since technological limitations prohibit sequencing an entire chromosome one base-pair at a time, current sequencing technologies involve breaking the DNA into many small fragments which then are partially sequenced. Various algorithms are then used to assemble these sequenced pieces of DNA \cite{pop2002,schatz2010}. In the past several years, researchers have progressed from sequencing small organisms such as the {\em Rhodobacter sphaeroides} with $4.42\times10^6$  base-pairs \cite{porter} to humans with $3.3\times10^9$ base-pairs \cite{lander2001} to even larger, more recent projects such as conifers with 2.4$\times10^{10}$ base-pairs \cite{pine}. As state of the art sequencing has changed, the computational challenges have grown immensely \cite{henson,schatz2010}. As a result, we need efficient, accurate computational approaches to problems that previously could be handled with relatively simple, easily implemented algorithms \cite{schatz2010}. There are a number of different steps in the genome sequencing process that could be examined and tested for improvement. These include identifying errors in the base-pair reads eg. \cite{li, trimble}, determining which pieces are repeat DNA eg. \cite{schaper,zerbino} or overlap eg. \cite{alba}, deciding the order in which pieces should be placed \cite{boetzer,dayarian,gritsenko}, or finally, our focus, determining the relative orientation of pieces in the assembly process \cite{huson, kececioglu}. 

 
We will provide a formal definition of the orientation problem in Section \ref{sec:definition} but first here is a simple analogy to keep in mind. Imagine opening a new jigsaw puzzle (representing DNA) that has an image on each side, call them side A and side B.  When you dump the puzzle out, in order to properly assemble it, the pieces must all have either the A side up or the B side up.  Sequencing an entire genome is like putting together a puzzle of the short sequenced reads.  Since DNA is double stranded, for each sequenced read, we have to figure out if it comes from the ``top'' or ``bottom'' strand.  Hence, the orientation problem in genome assembly is analogous to determining which side is up for all of the puzzle pieces.  If we have correctly fit together two pieces of the jigsaw puzzle, then we know that both pieces must have the same side facing up.  Similarly, sequencing data includes information (called linking-pairs, which we will describe later), that indicate the relative orientation between individual pieces.  In our approach to genome assembly, we encode this sequencing data into a network of interactions.   In this network, nodes are the sequenced pieces of DNA and each edge encodes information about the appropriate relative orientation for the node pair it connects.  

While the concept of orienting pieces is easily grasped, errors in real data can lead to a high level of conflicting information and turn solving the orientation problem into a hard computational challenge.  (The problem has been show to be NP-Complete \cite{huson,kececioglu}.)  Because of this, we must use a heuristic method to find an overall orientation.  Kececioglu and Myers \cite{kececioglu} have considered the orientation problem explicitly and have discussed various methods for solving it. In general, the orientation problem has been solved in conjunction with other components of the assembly process. In-fact, nearly all the other major methods we find (with the exception of Huson \textit{et al.} \cite{huson} who describe the method used in \textit{Celera}) describe their orientation algorithm as part of a larger scaffolding program \cite{boetzer, dayarian, gritsenko, pop2004}. Two methods of particular interest are used by \textit{Bambus} \cite{pop2004} and SOPRA \cite{dayarian}. \textit{Bambus}'s method, which we call ``Node-Centric Greedy'', is the most commonly used and what we will use as our baseline method. An explanation of both methods, and their limitations is presented in Section \ref{sec:others}.

\subsection*{Properties of partially processed genome sequence data}
\label{sec:datagen}

\begin{figure}[!h]
\begin{center}
\includegraphics[scale=.45]{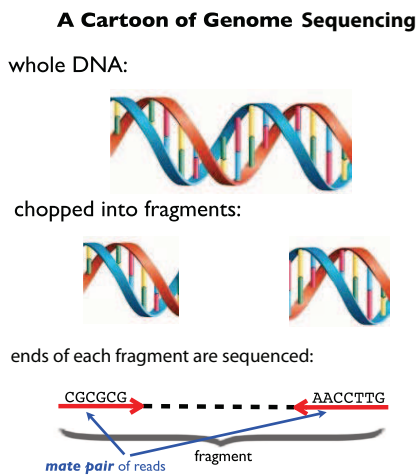}
\end{center}
\caption{An abstraction of the process of sequencing a genome: First the genome is split into fragments, and then the ends of the fragments are sequenced. These paired end sequences are called `mate-pairs'.}
\label{fig:cloning}
\end{figure}

In order to better understand the computation problem we face, let us examine the current DNA sequencing process. That is, the process by which a series of DNA base-pairs are determined.   The leading approach for generating genome sequence data is Whole Genome Shotgun sequencing (WGS). Our explanation here is a paraphrase of material from several review articles: \cite{henson, pop2002, schatz2010}. In WGS, many nearly identical copies of DNA from a large number of cells get shredded randomly into fragments of 200-20000 bases long. The fragments are then size selected to obtain a libraries with known size means and standard deviations. Then 100-400 bases on both ends of the fragments are sequenced, forming the basic data unit\--a mate pair of reads. This process is shown in Figure \ref{fig:cloning}. Note that the mate-pairs are generated by reading inward from both ends of each DNA fragment, thereby having an inherent relative orientation.

\begin{figure}[!h]
\begin{center}
\includegraphics[scale=.45]{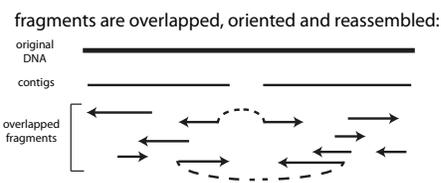}
\end{center}
\small\normalsize
\caption{Sequenced fragments are overlapped, and reassembled into 'contigs', then full genomes. The mate-pair information is vital in this reassembly process to determine orientation and placement.}
\label{fig:assembly}
\end{figure}

By design, we know the mean base-pair length and the standard deviation for the fragment distribution as well as the mutual orientation of mate pairs. The first computational step in the assembly process is to build larger contiguous segments (contigs) from the read sequences by overlapping and combining fragments as shown in Figure \ref{fig:assembly}. The process of building contigs is usually followed by a `scaffolding' process in which the contigs are ordered and oriented into larger components (called scaffolds). During scaffolding, the mate-pairs for which the two mates ended up in two different contigs are vital to the reassembly process as they are used to determine the correct order and orientation for the collection of contigs.  We call such mate-pairs linking mates.  A linking mate-pair specifies the relative orientation and approximate relative position of two contigs. A set of contigs connected by linking mates can be used to form a contig network.

\begin{figure}[!h]
\begin{center}
\includegraphics[scale=.6]{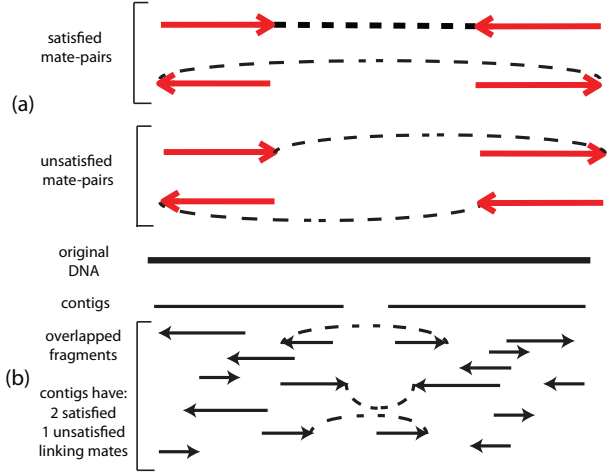}
\end{center}
\small\normalsize
\caption{(a) Shows two pairs of examples. The top pair show two ways reads can be oriented (opposite directions) so that the mate-pair is satisfied. The bottom pair are two ways that reads can be oriented (same direction) which makes them unsatisfied.\\
(b) Shows how satisfied and unsatisfied mate-pairs combine to form conflicting data between contigs.}
\label{fig:matepairs}
\end{figure}

When mate-pair reads are originally generated the two reads have opposite orientations since they are read from opposing directions on the same fragment of DNA (as shown in the bottom of Figure \ref{fig:cloning}). A linking mate is therefore satisfied if the orientations of the two contigs where the reads are placed and the orientations of the two mates within these contigs imply that the linking mates are oppositely oriented.  A linking mate is unsatisfied if the implied orientations of the two reads are the same. These two situations are shown in Figure \ref{fig:matepairs} (a). In general some linking mates could be conflicting due to errors in generating or reporting the mates, repeated genomic sequence that may lead to incorrect read placement, etc. This conflict from a combination of the satisfied and unsatisfied links occurring in an assembled contig is shown in Figure \ref{fig:matepairs}(b). We informally define the orientation problem as finding an orientation for each contig that minimizes the total number of unsatisfied mate-pairs. A formal definition is found in Section \ref{sec:definition}.

The remainder of the paper is structured as follows: Section \ref{sec:results4} compares our new algorithm (Hierarchical Greedy) to Node-Centic Greedy on both experimental data (\textit{Rhodobacter sphaeroides bacteria}) and generated faux data.   Our methods are described in Section \ref{sec:method}, beginning with a formal definition of the orientation problem and followed by a description the basic structure of our algorithm and an elaboration of the algorithms details. Finally, in Section \ref{sec:discussion} we provide a detailed description of two other existing methods, and their comparative limitations, followed by a short summary.

\section{Results}
\label{sec:results4}
In this section we report the performance of our method on two sets of data. First we apply the method to contigs of Rhodobacter sphaeroides bacteria, produced by the MSR-CA assembler and show that the resulting orientation agrees with the finished sequence for that bacteria. We then use a faux genome assembly we generated for which the correct answer is known before any errors are introduced. We study the performance and stability of our algorithm on this faux assembly and compare its pefromance to the Node-Centric Greedy method as we vary the initial conditions (initial designations for the contig orientations) and introduce noise (conflicting mate pairs) into the generated data.

\subsection*{\textit{Rhodobacter Sphaeroides bacteria}}
We used the MSR-CA assembler version 1.8.3 to produce contigs and scaffolds from Illumina data for Rhodobacter sphaeroides bacteria. The data has been downloaded from the NCBI Short Read Archive. The data consisted of (1) a paired end library (PE), in which reads were generated from both ends of 180-bp DNA fragments (SRA accession SRR081522); and (2) a ``jumping'' library (SJ) in which paired ends were sequenced from 3600-bp fragments, (SRA accession SRR034528). We down-sampled both libraries to 45x genome coverage. The available finished sequence for the organism allowed us to evaluate the correctness of our orientation solution. 

Since the MSR-CA assembler reports read positions in the final assembled contigs it was easy to convert the contigs into a graph. We ignored single (weight 1) mate pair links between the contigs, and excluded mate-pairs that are interior to the contigs or whose read placements indicate that they cannot plausibly (within five standard deviations) link contigs in a non-overlapping fashion. We performed our orientation studies on the biggest chromosome of the \textit{R. sphaeroides}, 3.2Mb long.  The MSR-CA assembler created a single scaffold for that chromosome and, according to the mapping of contigs in the scaffold to the finished sequence, all contigs in the scaffold were assembled and oriented correctly.

The scaffold that we were using contained 260 contigs connected with 8829 mate pair links. This is approximately 34 mate-pairs per contig. This is a real sequencing data set and it is expected that some portion of the data is in chimeric or misoriented pairs, and the orientation algorithm must have enough skill to correctly resolve the conflicting data.

\begin{figure}[!b]
\centering \includegraphics[scale=.65]{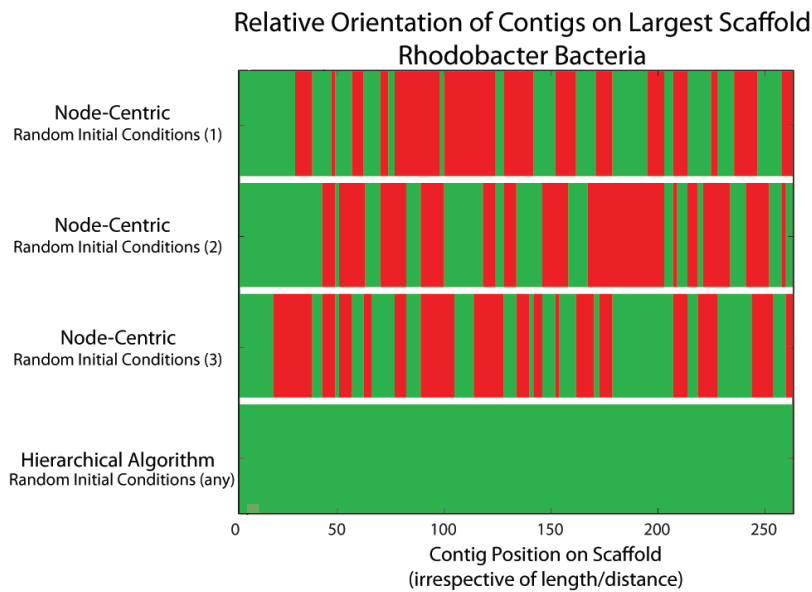}
\caption{Results of applying Node-Centric Greedy and Hierarchical Greedy to \textit{R. sphaeroides} assembly data. Orientations are plotted as a function of the known contig position on the scaffold, with red/green represent orientations of -1 \& +1 respectively.  The top three rows show results from using the Node-Centric Greedy algorithm starting from three different random initial conditions.  Note the high variance in the solutions. The bottom row shows results of the Hierarchical method which are the same for any initial condition.  Several contigs in between positions 1 and 50 had no orientation information available and are not shown.}
\label{fig:bar}
\vspace{-20pt}
\end{figure}

Using this data set of contigs from the primary scaffold we performed two experiments. We first used data from \textit{pre}-Celera orientation and scaffolding as initial conditions to compute orientations using the Node-Centric Greedy and Hierarchical Greedy methods. Both methods achieved a solution in complete agreement with the known solution and did correct several initially incorrect contig orientations. The second experiment, shown in Figure \ref{fig:bar}, was to randomize the initial contig orientations from a correct solution. With this experiment, the Node-Centric Greedy algorithm is unable to find the correct solution. Figure \ref{fig:bar} shows three examples of final solutions from the Node-Centric Greedy and one from Hierarchical Greedy.

A horizontal plot that matches the known solution would have all green (all same) or red (all reversed, matching within an overall flip). A few contigs (in positions between 1 and 50) had no mate-pair information available, and therefore are not shown. All randomized initial conditions led to the same correct solution for our Hierarchical Greedy method, while the Node-Centric Greedy method produced a variety of different solutions, all incorrect (three of which are shown as an example).

\subsection*{Faux data}
\begin{figure}[!hb]
\begin{center}
\includegraphics[scale=1.05]{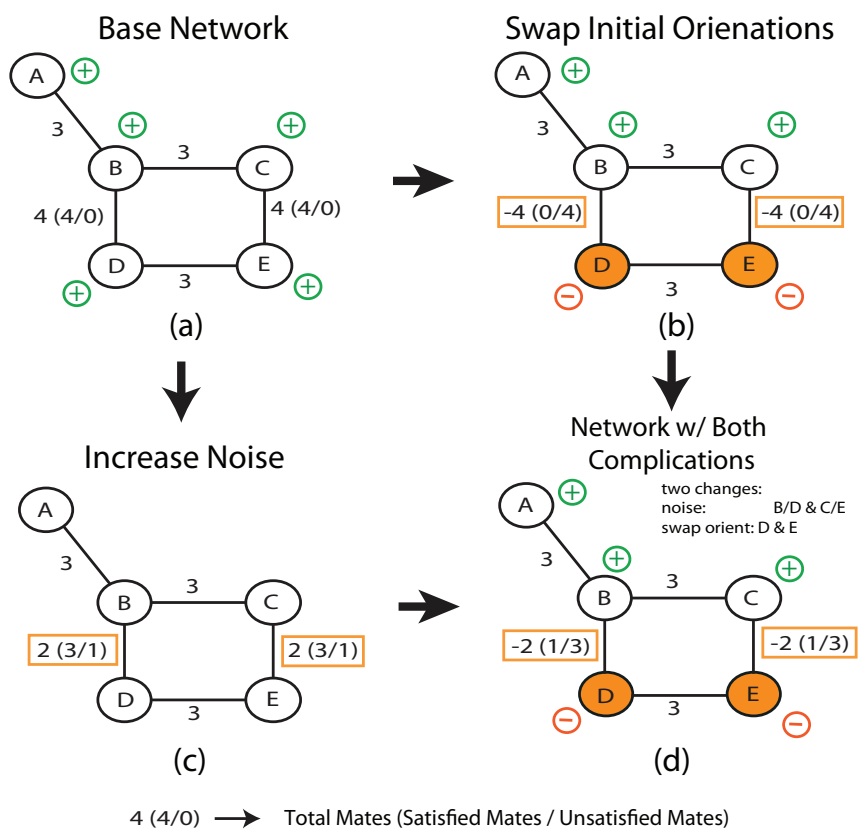}
\end{center}
\caption{Examples of errors introduced in faux data.  Orientations are indicated by green pluses or orange minuses.  Edge values report the number of satisfied - unsatisfied mate pairs between the contigs.  Satisfied/Unsatisfied counts are given in parenthesis.
(a) Shows a base network with no conflicting links in which all contigs are properly oriented.
(b) Shows how swapping an initial orientation of node D changes the configuration from the base network.
(c) Shows how introducing noise on edges $B\leftrightarrow D$ and $C\leftrightarrow E$ changes the configuration from the base network.
(d) Shows how a more complex input configuration can be generated by both varying initial orientations and introducing noise (used later in Section \ref{sec:discussion}). }
\label{fig:exerrors}
\end{figure}

We produced a faux contig (and related mate-pair) data set using the same base length and library distributions for the constructing fragments as those in \textit{Rhodobacter sphaeroides bacteria}. In our base network, all contigs were correctly oriented in the 'positive' direction and the mate-pair data did not contain any conflicting mate-pairs.  Hence, if the base network is used as input to either Node-Centric Greedy or Hierarchical Greedy, the output orientations of all the contigs remains unchanged, since the system is already correctly oriented.  Starting from the base network, we modified the input to our algorithm in two ways, 1) by varying the initial conditions (initial orientations of the contigs) and 2) introducing noise (conflicting links) into the data. These are demonstrated in Figure \ref{fig:exerrors}. 

In order to introduce modifications in the initial conditions we start with the base network configuration and randomly flip the orientation of some fraction of the contigs. This initial change in orientation forces all the mate-pairs connecting the contig to its neighbors to become unsatisfied. This means that a final (correct) solution should still have zero unsatisfied mate-pairs, and the correct solution should indicates that any initially flipped contig should be flipped again (returned to the 'positive' orientation). This type of modification is shown in Figure \ref{fig:exerrors}(b). 

For the other modification to the dataset we introduce noise by switching the orientations for one (or more) of the reads in some linking mate-pairs. This creates a subset of mate-pairs which have conflicts. The conflicts make a perfect solution no longer possible in most cases as shown in Figure \ref{fig:exerrors}(c). For the results shown below we introduced this noise randomly into each of the 20 trials, so the best possible solution, in terms of satisfied mate-pairs, does vary. 

When these modifications are combined as in \ref{fig:exerrors}(d), we get more complex input configurations which cause problems in simple orientation schemes (see Section \ref{sec:discussion} for more details). To summarize: any orientation algorithm that is working well should produce an optimal solution regardless of any changes in the initial conditions from the base configuration (since we have not introduced conflicting data). However, when we introduce noise, the original solution may no longer be optimal and we should expect that a different solution from the original may better satisfy the new mate pair data.

\begin{figure}[!h]
	\begin{center}
	\begin{subfigure}{.9\linewidth}
		\centering \includegraphics[scale=.7]{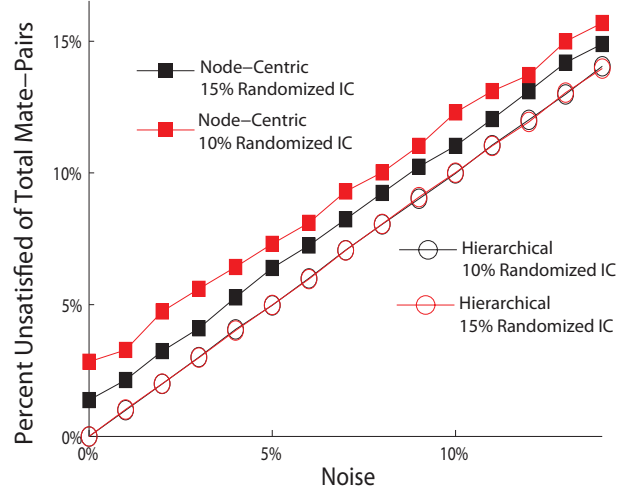}
		\caption{Noise vs. unsatisfied mate-pairs for node-centric and hierarchical methods}
		\label{fig:noisebad}
	\end{subfigure}
	
	\begin{subfigure}{.9\linewidth}
		\centering \includegraphics[scale=.6]{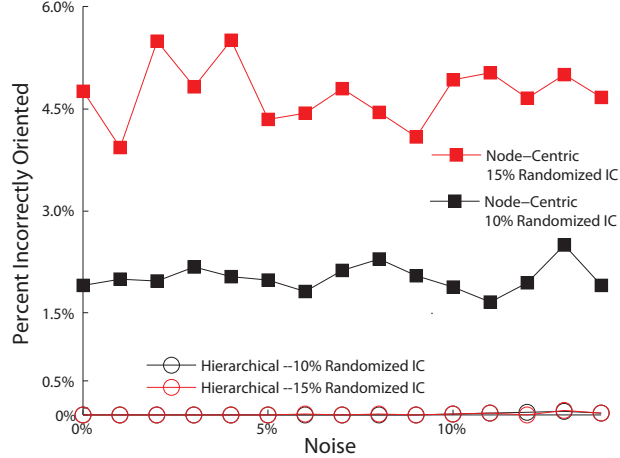}
		\caption{Noise vs. incorrectly oriented mate-pairs for node-centric and hierarchical methods}
		\label{fig:noisewrong}
	\end{subfigure}
	\end{center}
\small\normalsize
\vspace{-20pt}
\caption{In both (a) \& (b) two pairs of lines are shown. The solid, square points show average solutions for the Node-Centric Greedy algorithm. The empty circles show average solutions for the Hierarchical Greedy method. The two colors (black \& red) represent starting with 10\% and 15\% randomized initial conditions (like (b) in Fig \ref{fig:exerrors}). The \textsl{x}-axis shows increasing noise (from (c) in Fig \ref{fig:exerrors}). In (a) the \textsl{y}-axis is the percent of unsatisfied mate-pairs. In (b) the \textsl{y}-axis is the percent incorrectly oriented compared to the base, expected solution. }
\label{fig:fauxdata}
\end{figure}

Figure \ref{fig:fauxdata} shows a comparison between the Hierarchical Greedy (open circles) and Node-Centric Greedy (filled squares) algorithms. The colors differentiate between 10\% differance (black) and 15\% difference (red) from perfect initial conditions. For both (a) and (b) the \textsl{x}-axis shows an increasing noise rate (conflicting information) in the data, i.e. more mate-pairs which become impossible to satisfy. Figure \ref{fig:fauxdata}(a) shows the count of unsatisfied mate-pairs on the \textsl{y}-axis, for which Hierarchical Greedy always ties or out-performs the Node-Centric Method. On this figure, we would expect a straight line with slope=1 through the origin if the method is correcting all possible swapped orientations ([c] in Figure \ref{fig:exerrors}). We can see that the Hierarchical Greedy method achieves this, while the Node-Centric Greedy method does not. Node-Centric Greedy does achieve a slope of approximately 1, however it does not correctly orient all the mate-pairs even with 0\% noise. Figure \ref{fig:fauxdata}(b) on the \textsl{y}-axis shows the percentage of incorrectly oriented contigs in the final solution, based on the original base solution.
\newpage
\noindent
\newpage

\section{Materials and Methods}
\label{sec:method}
\subsection*{The orientation problem}
\label{sec:definition}
\noindent
Before describing our new algorithm, we formally define our problem:

Given a collection of contigs ($C$) and the satisfied ($S^0$) and unsatisfied ($U^0$) mate-pairs connecting them given an initial orientation, define a coupling strength between two contigs \textit{i} and \textit{j}:\\
\begin{equation}
c_{ij}=f(S^0_{ij},U^0_{ij})
\end{equation}
Further, we denote the orientation of a contig $i$ as $\sigma_i$ and use the values of $\pm 1$ to represent the two possible orientations. Without loss of generality, we set the initial orientation (when the coupling strength is defined) of all contigs  to $+1$. The orientation problem then is to identify the set of orientations that maximize the sum:
\begin{equation}
S=\sum\limits_{i,j\in C}{c_{ij}\sigma_i \sigma_j }
\label{eq:sumC}
\end{equation}

Note that the coupling strengths are fixed (after contig assembly) according to our definition above. Also, while we have indicated the coupling strength should be a function of satisfied and unsatisfied mate-pairs, there could be additional genomic information that affects orientations, e.g. mate-pairs initially deemed unreliable. Such additional information could be incorporated through appropriate modifications to the $c_{ij}$.

In order to better illustrate how the sum defined by Eq. \eqref{eq:sumC} may be increased by flipping the orientations of some of the contigs, we find it useful to explicitly define a simple $c_{ij}$ as $S^0_{ij}-U^0_{ij}$.  We can then write the terms $c_{ij}\sigma_i \sigma_j $ as:
\begin{equation}
c_{ij} \sigma_i \sigma_j=(S^0_{ij}-U^0_{ij})\sigma_i \sigma_j = S_{ij}(\sigma_i,\sigma_j)-U_{ij}(\sigma_i,\sigma_j)
\label{eq:xij}
\end{equation}
Here, for the specified orientations $\sigma_i$ and $\sigma_j$, the term $S_{ij}(\sigma_i,\sigma_j)$ gives the number of satisfied mate pairs connecting contigs $i$ and $j$ and $U_{ij}(\sigma_i,\sigma_j)$ gives the number of unsatisfied mate pairs connecting them. 
Ideally a final orientation solution will have every $c_{ij}\sigma_i \sigma_j$ term positive. This would occur either by having a positive $c_{ij}$ and matching $\sigma_{i,j}$ (both $+1$ or $-1$) \textit{or} by having a negative $c_{ij}$ and different $\sigma_{i,j}$. In $c_{ij} \sigma_i \sigma_j$ reversing a contig's (e.g. $j$) orientation would change the sign of $\sigma$ (e.g. $\sigma_j$) and thereby change the sign on both terms on the right hand side in Equation \eqref{eq:xij}.  In other words, reversing a contig's orientation makes all the unsatisfied mate-pairs connected to it become satisfied and all the satisfied mate-pairs connected to it become unsatisfied.  Thus, we can say that finding the maximum of the sum in Eq. \eqref{eq:sumC} above is equivalent to finding the set of orientations that minimizes the number of unsatisfied mate pairs.

The notation here is reminiscent the spin-glass problem in physics.   Alternative formulations have been presented by Huson \textit{et al.} \cite{huson} and Kececioglu and Myers \cite{kececioglu}.  Both sets of authors map their definitions to NP-Complete problems.  Also, Delorme and Poljak have shown a version of the Max-Cut problem (a traditional NP-complete problem) analogous to our formulation \cite{delorme}. 

\subsection*{Basic Algorithm}
\label{sec:basicalg}
Our new algorithm for solving the orientation problem is based on the process of hierarchical clustering. Hierarchical clustering was originally proposed and used for phylogenetic studies \cite{sokal} and now related methods are utilized in many different areas such as information retrieval, multi-variate data analysis and community finding in networks \cite{steinbach, rizzo, shen}. The central idea is to build a tree-like structure, called a dendrogram, which is a meaningful ordering of merges for the nodes that creates progressively larger and larger sets (clusters) upon which one can continue to make merging decisions \cite{ward}. 

Given a matrix of interaction data, there are two important elements to finding an ordering of merges, a weighting metric for determining relationships between single nodes and a {\em re}-weighting scheme for computing weights between merged clusters. Traditional hierarchical clustering approaches require that all weights are non-negative, allowing various agglomerative re-weighting schemes.  In these schemes, one can compute the weights between cluster pairs independent of the order of prior merges.  Here, we wish to introduce a variant of hierarchical clustering, in which the nodes have signs (orientations) and the weights between node clusters depend on previous join steps and orientation decisions.
This prevents us from using standard re-weighting schemes, such as average linkage clustering \cite{murtagh, sokal} directly. Instead we introduce a combined weighting, orienting, and re-weighting scheme.\\

\noindent
The rationale motivating our weighting scheme is as follows:
\begin{itemize}
\item If there is no conflicting information for a node pair, the metric should be proportional to the number of the mate pairs connecting the pair. 
\item If a node pair has conflicting mate pair information, the link weight for the node pair should be reduced (indicating that the orientation information is not completely trustworthy).
\item Conflicts leading to equal number of satisfied and unsatisfied mate-pairs should produce zero weight (no connection).
\item Link weights between clusters of contigs should be comparable in scale to link weights between individual contigs.
\end{itemize}
\noindent
With these in mind, we propose the following method for determining the link weight between contig clusters:\\
\vspace{.1cm}

Given two clusters of contigs $A$ and $B$, and the orientations of the nodes within them, we define the ``net between-cluster edge satisfaction'':
\begin{equation}
\Delta_{AB}:= \sum\limits_{i\in A, j\in B} (S^0_{ij}-U^0_{ij})\sigma_i \sigma_j=\sum\limits_{i\in A, j\in B}S_{ij}-U_{ij}=S_{AB}-U_{AB}
\label{eq:edgesat}
\end{equation}
Define the ``total mate-pair count'' (which does not depend on orientations):
\begin{equation}
\tau_{AB}:= \sum\limits_{i\in A, j\in B} (S_{ij}+U_{ij}) = S_{AB}+U_{AB}
\end{equation}
Finally, define a scaling factor:
\begin{equation}
f_{AB}:=\frac{\left| \Delta_{AB}\right|}{\tau_{AB}}
\end{equation}
Then the weight on an edge is the magnitude of the difference scaled by this factor, divided by the number of elements in $A$ and $B$:
\begin{equation}
w_{AB} := \frac{1}{\left|A\right|\left|B\right|} 
\left| \Delta_{AB} \right| 
f_{AB} = \frac{1}{\left|A\right|\left|B\right|} 
 \frac{ \left(S_{AB}- U_{AB}\right)^2 }{ S_{AB}+U_{AB}} 
\label{eq:wght}
\end{equation}
\label{def:wght}

Note that $w_{AB}$ is computed every time we create a new cluster (by merging singleton nodes, or previous clusters), thus defining the `new' weight from that cluster to every other cluster.
In the following discussions we will use the term `cluster' to refer to both unmerged (singleton) and combined contigs. \\

\noindent
Having defined a weighting metric and a re-weighting scheme our algorithm is as follows:\\
\newline
\vspace{.1cm}
\begin{minipage}{\linewidth}
\textbf{Hierarchical Greedy Algorithm}
\begin{enumerate}
\item Choose the edge with the maximum weight* using the given metric

 \hspace{0.75cm} and identify the two clusters on either side, call them clusters $A$ and $B$
\item Determine if either cluster should be flipped.\\
\indent If yes (i.e if the number of unsatisfied edges, $U_{AB}$, is greater than the number of satisfied edges, $S_{AB}$:
\begin{itemize}
\item \small{Determine the best cluster to flip, based on the effect on the remaining network. (Choose the cluster that results in the fewest unsatisfied edges between $AB$ and the other nodes in the network.)}
\item \small{Flip the identified cluster (i.e. flip the orientations of all nodes in the cluster). }
\end{itemize}
\item Merge the two clusters into a new cluster.
\item Zero out the edge chosen for merging (now an internal edge)
\item Recompute the edge weights connecting the new cluster to every other cluster according to the weighting metric.
\item Repeat steps 1-5 until no edges remain (this may leave more than one cluster, if some clusters have no edges connecting them).
\end{enumerate}
\end{minipage}
\vspace{.2cm}* See the following section for a more in-depth discussion of this choice in the case of ties.
\vspace{.4cm}

\begin{figure}[!h]
\begin{center}
\includegraphics[scale=.75]{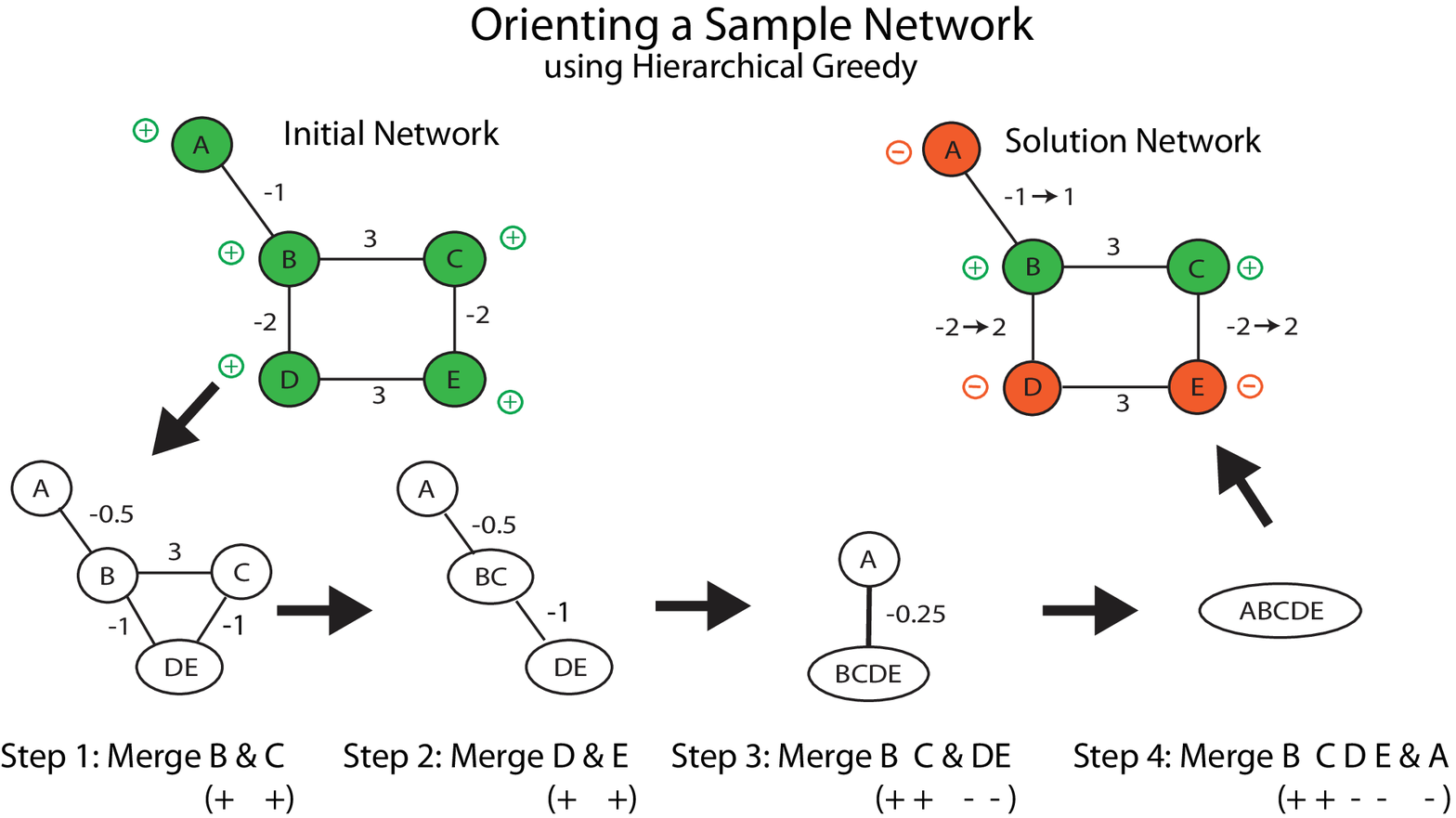}
\end{center}
\caption{This figure illustrates how our hierarchical method orients a network that causes trouble for Node-Centric Greedy. 
We start with an all $+$ orientations (upper left), showing each merge and redefinition of weights until we have merged all the nodes together (lower right). This gives us a final solution shown in the upper right.  Note: To improve readability in the figure the edge weights we show use $\frac{\Delta_{AB}}{|A||B|}$ rather than Eq. \eqref{eq:wght}, however the join order would be the same if Eq. \eqref{eq:wght} were used instead. Negative signs indicate that the number of unsatisfied mate pairs is greater than the number of satisfied mate pairs for that configuration. }
\label{fig:hiermerge}
\end{figure}

As stated in step 6, the merging process is repeated until each connected component in the network has been merged into a single cluster. Figure \ref{fig:hiermerge} illustrates the step-by-step application of our method to the small sample network for which the Node-Centric Greedy method fails (see Section \ref{sec:discussion}).

This iterative process of merging and reorienting is where the novelty of our procedure arises. By merging previously oriented elements and revising their weights relative to other nodes (which depends on the orientations), it becomes impossible to precompute all the joins as in standard hierarchical clustering methods. Furthermore, the merging into orientable groups introduces an important and novel approach to genome orientation. The merging forces the algorithm to fix relative orientations between elements, maintaining correctness while allowing any connection which has not been evaluated to be correctly oriented later.

\subsection*{Algorithm details}
\label{sec:algdetails}
\begin{figure}[!h]
\begin{center}
\includegraphics[scale=1]{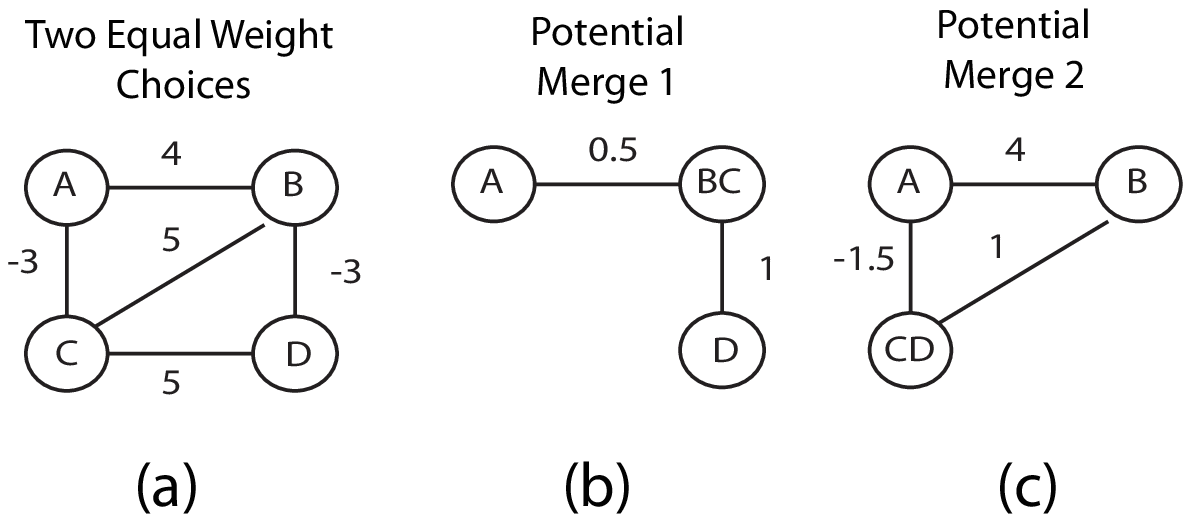}
\end{center}
\caption{This figure shows a network (a) that is challenging for Hierarchical Greedy to merge properly if no method for breaking ties between equally weighted links is used. Two different potential merges (b \& c) are shown. (b) is worse because it loses more potential edge information by combining in two sets of negative and positive edges while (c) only merges one negative and positive edge.}
\label{fig:equalex}
\end{figure}

Due to the nature of sequencing and contig assembly data we are not working with a broad distribution of edge weights in the network. This means that in many cases, step 1 of the algorithm will actually find two (or more) edges with equal weights to merge. An example of this situation can be seen in Figure \ref{fig:equalex}(a), where both edge B$\rightarrow$C and C$\rightarrow$D have weight 5. The two potential merges are given in parts (b) and (c) of the figure.

A simple initial solution to this issue is to randomly choose among all equally weighted edges at any iteration. This can produce a distribution of final solutions. An alternative is to perform all equally weighted merges in a parallel fashion. We have found this second option to actually produce solutions that are significantly worse (with a higher unsatisfied edge count) compared to the random choice method. Because we are interested in stable and accurate performance of our algorithm, we want an approrpriate {\em deterministic} method for breaking ties between equally weighted edges.   To achieve this, we introduce a 'Maximum Options Remaining' (MOR) measure.  

The MOR measure tracks the number of unsatisfied mate pairs that remain unfixed.  Once two nodes are placed in the same cluster, their relative orientations are fixed and any unsatisfied mate pairs between them can never become satisfied in subsequent merges.  Using the MOR principle, we delay fixing mate pairs as unsatisfied until further information has been incorporated into the orientation choices through subsequent merges. For example, applying the the MOR principle in Figure \ref{fig:equalex} we would choose the merge shown in (c). We can see that in part (c) fewer unsatisfied edges (one, $B\leftrightarrow D$, instead of two, $B\leftrightarrow D$ and $A\leftrightarrow C$) have been merged into other edges. This also has the effect of maintaining the largest absolute weight on the edges, indicating that further choices will be able to make larger changes.

We can express the MOR principle mathematically in the following way. Given a potential merge, $AB$, in which $A$ and $B$ are sets of contigs, along with the orientations that would be set by that merge, we denote the set of shared neighbors between $A$ and $B$ as $X$.  We define $R$, `options remaining', as:\\
\begin{align}
R_{AB}= \left|S_{AX}-U_{AX}+S_{BX}-U_{BX} \right|
\end{align}
Then when comparing two merges $AB$ and $CD$, $AB$ is chosen if:\\
\begin{equation}
R_{AB}>R_{CD}
\end{equation}
Otherwise, merge $CD$ is chosen.

Utilizing the MOR principle, nearly all equal choices can be given a relative ordering. While it is still possible to find two edges with equal weight and equal options remaining, the occurrences are far less frequent. This low collision rate occurs due to the sparse nature of our network, and, under testing, arbitrarily choosing between them has not shown any affect on the final solution.

\section{Discussion}
\label{sec:discussion}
\subsection*{Breakdowns in existing algorithms}
\label{sec:others}
In the introduction, we mentioned several other existing algorithms, most notably the Node-Centric Greedy from \textit{Bambus} \cite{pop2004} and the one used by SOPRA \cite{dayarian}. In this part of the discussion section, we will give more details about these methods, highlighting problems that Hierarchical Greedy can overcome. We begin with details on the Node-Centric Greedy method and several examples, then discuss SOPRA's method.

\noindent
The Node-Centric Greedy method works on a contig network in the following way:\\
\vspace{0.2cm}\\
\noindent
\begin{minipage}{\linewidth}
\begin{enumerate}
\item Sum $S_{ij}(\sigma_i,\sigma_j) - U_{ij}(\sigma_i,\sigma_j)$ for each $i\in C$ (node in graph)
\item Choose the most negative (most unsatisfied) {\em{contig}}.
\item Reorient that contig (and fix it).
\item Recalculate node-sums of all connected nodes.
\item If any node has a negative sum return to step 2
\end{enumerate}
\end{minipage}
\newline
\vspace{.4cm}
\newline
To elaborate a little, step one adds up the information from all of the mate-pairs that are incident on a contig. From the definition of the orientation problem in Section \ref{sec:definition}, this corresponds to finding all the $z_{i}$ where $z_{i}=\sum\limits_{j \in C}{c_{i j } \sigma_{i} \sigma_j }$. By keeping a master-list of all these sums, the algorithm can pick the most unsatisfied contig (the one with the most negative $z$ value) and reorient that contig. This has a trickle down effect of changing the contig sum for any connected contigs. 

\begin{figure}[!h]
\begin{center}
\includegraphics[scale=1]{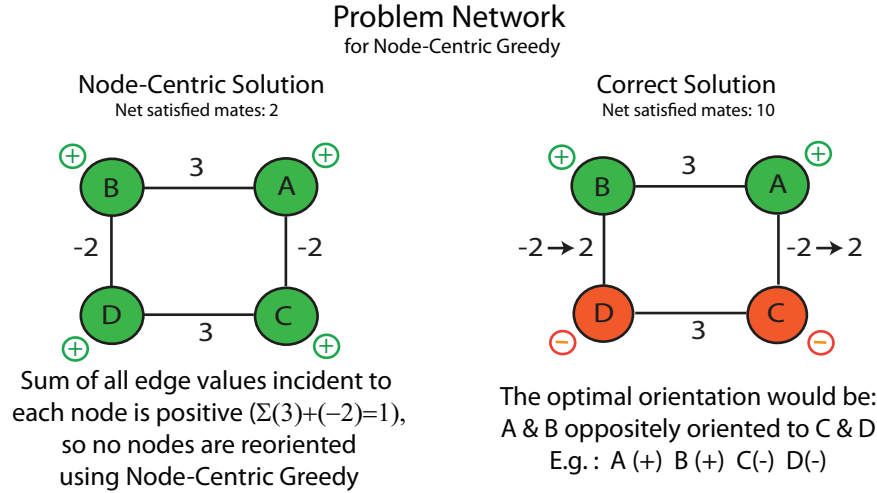}
\end{center}
\caption{This figure shows two versions of a small graph. Values on the edges represent the $c_{ij}$ as defined from Eqn. \eqref{eq:xij}. 
The left version demonstrates how the Node-Centric Greedy will not impose any orientation change while the right shows an optimal final orientation.}
\label{fig:breakgrdy}
\end{figure}

Overall this method is fairly effective, fixing the biggest problems first, and continuing to fix problems until there do not appear to be any problems remaining. However, the method suffers from the flaw of data agglomeration.  What we mean is that by examining the sum, and only orienting those contigs with a negative sum, it becomes possible to `bury' some of the information in unsatisfied mate pairs. The small network shown in Figure \ref{fig:breakgrdy} illustrates this point, and shows a case in which Node-Centric Greedy can return an incorrect orientation. 

\begin{figure}[!h]
\begin{center}
\includegraphics[scale=1]{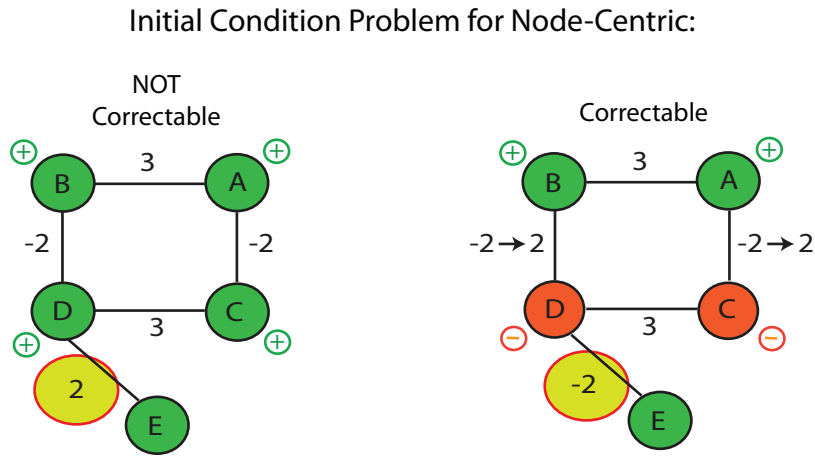}
\end{center}
\caption{This figure shows two versions of a small graph. Values on the edges represent the $c_{ij} \sigma_i \sigma_j$ from Eqn. \eqref{eq:xij}. 
In the left version, Node-Centric Greedy will not impose any orientation change, while the right shows a change in initial conditions that allows Node-Centric Greedy to find a correct final orientation.}
\label{fig:breakgrdy_IC}
\end{figure}

Another issue with Node-Centric Greedy is its sensitivity to initial conditions. We already demonstrated in the results section how robust our new algorithm is, but a simple addition of a spur to the network in Figure \ref{fig:breakgrdy} can demonstrate this. The modified network is shown in Figure \ref{fig:breakgrdy_IC}. When the spur between node D \& E is positive (left), the nodes maintain a positive edge-sum just like in the previous graph. However, changing the initial condition to be negative over that edge produces a negative node-sum on node D, and allows the entire network to be solved correctly (it breaks the symmetry). Since this edge change could be produced by a simple changing of node E's initial condition producing a disturbing initial conditions problem. Both graphs permit a solution with all positive mate-pairs.

While the node-centric method is demonstratively flawed, the method presented in by Dayarian {\em et al.} in \cite{dayarian} is a bit more robust. The method in \cite{dayarian} is actually two different ideas combined. The first part of the method finds articulation points that break the network into smaller components. By breaking the network into components in this manner they produce several significantly smaller sub-graphs (see \cite{ausiello, hopcroft, khuller} and others for details on articulation points and components). Because these sub-graphs generally have a much smaller number of nodes and edges, solving the orientation problem exactly on these sub-graphs can often be accomplished quickly even though the sub-problems remain NP hard (because N is small). It can be proven mathematically that if the sub-parts can be solved exactly, then breaking the problem into pieces that are then subsequently reassembled does not produce a worse orientation than solving the whole problem directly, and the reassembly process can be done in linear time. This means that if the contig network can be broken completely into small, quickly solvable pieces the overall orientation process will be significantly faster than if the algorithm used does not break the problem into sub-parts. 

The second part of their method, however, leaves significant room for improvement. While breaking the network allows most subgraphs to be solved exactly, some are still too large, therefore Dayarian {\em et al.} utilize a heuristic method to solve them. They chose to implement a standard Ising model approach to properly orient their remaining large graphs. As the authors themselves state, the Ising model approach they use may not give optimal solutions if ``there are highly-connected components of moderate or large size'' \cite{dayarian}. Other investigations (not detailed here) have shown that mammalian genomes exhibit exactly this trait. 

\begin{figure}[!ht]
\begin{center}
\includegraphics[scale=1]{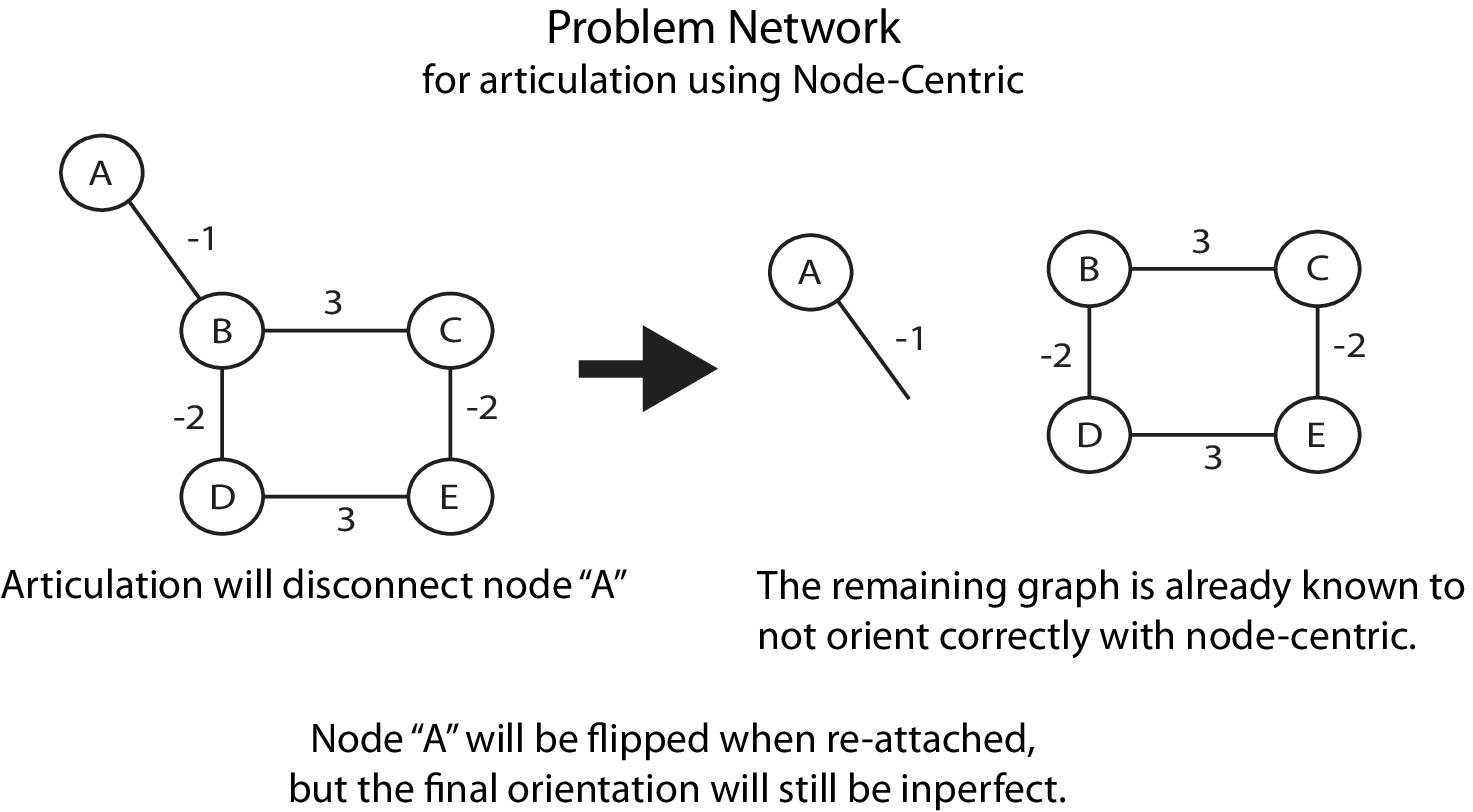}
\end{center}
\caption{This figure shows a 5-node network that the articulation method can break apart (left breaks into the right). However, the resulting sub-networks will not be oriented correctly if the node-centric algorithm is used as the component-solving method. This means when recombined, even though node A will be reoriented, the whole network will be non-optimally oriented.}
\label{fig:articulation}
\end{figure}

Basically, their approach to problem reduction is fully dependent on the quality of the actual method to find orientations in the subgraphs. As a demonstration, we present a slightly modified version of Figure \ref{fig:breakgrdy} that shows one additional node added on. If we perform articulation on the network (which, in this case, corresponds to breaking that node back off), and we then use Node-Centric Greedy as our component-solving method, we are again left with an incorrect solution. Figure \ref{fig:articulation} shows this process.

\subsection*{Summary}
\label{sec:summary}
Our results clearly show that a Node-Centric Greedy algorithm performs poorly for both random initial conditions and noisy data. Our new approach provides several advantages that can address these concerns. First, by focusing on edges (mate-pairs), we can capitalize on the fact that some areas of the genome are easier to sequence correctly. These correctly sequenced regions will have a high agreement, and high coverage. Second, by joining based on edges we only lock in correct relative solutions, rather than a potential mix of good and bad solutions. Finally, our technique can be incorporated into any assembly program to provide partial or complete solutions to the orientation problem at several different stages since it does not depend on placement of contigs in a scaffold or contig assembly (if applied to orienting read fragments) to produce an orientation.

\section*{Acknowledgments}
This project was supported in part by National Research Initiative Competitive grants 2009-35205-05209 and 2008-04049 from the United States Department of Agriculture National Institute of Food and Agriculture. This research was also supported by National Institutes of Health grants R01-HG002945 and R01-HG006677.

\clearpage

\bibliography{gabib,graphbib,cluster,introbib}

\begin{thebibliography}{10}
\providecommand{\url}[1]{\texttt{#1}}
\providecommand{\urlprefix}{URL }
\expandafter\ifx\csname urlstyle\endcsname\relax
  \providecommand{\doi}[1]{doi:\discretionary{}{}{}#1}\else
  \providecommand{\doi}{doi:\discretionary{}{}{}\begingroup
  \urlstyle{rm}\Url}\fi
\providecommand{\bibAnnoteFile}[1]{%
  \IfFileExists{#1}{\begin{quotation}\noindent\textsc{Key:} #1\\
  \textsc{Annotation:}\ \input{#1}\end{quotation}}{}}
\providecommand{\bibAnnote}[2]{%
  \begin{quotation}\noindent\textsc{Key:} #1\\
  \textsc{Annotation:}\ #2\end{quotation}}
\providecommand{\eprint}[2][]{\url{#2}}

\bibitem{huangmurray}
Huang KG, Murray FE (2010) Entrepreneurial experiments in science policy:
  Analyzing the human genome project.
\newblock Research Policy 39: 567--582.
\bibAnnoteFile{huangmurray}

\bibitem{lander2001}
Lander ES, Linton LM, Birren B, Nusbaum C, Zody MC, et~al. (2001) Initial
  sequencing and analysis of the human genome.
\newblock Nature 409: 860–921.
\bibAnnoteFile{lander2001}

\bibitem{lander2011}
Lander ES (2011) Initial impact of the sequencing of the human genome.
\newblock Nature 470: 187--197.
\bibAnnoteFile{lander2011}

\bibitem{pop2002}
Pop M, Salzberg S, Shumway M (2002) Genome sequence assembly: Algorithms and
  issues.
\newblock Computer : 47–54.
\bibAnnoteFile{pop2002}

\bibitem{schatz2010}
Schatz MC, Delcher AL, Salzberg SL (2010) Assembly of large genomes using
  second-generation sequencing.
\newblock Genome Research 20: 1165 --1173.
\bibAnnoteFile{schatz2010}

\bibitem{porter}
Porter SL, Wilkinson DA, Byles ED, Wadhams GH, Taylor S, et~al. (2011) Genome
  sequence of rhodobacter sphaeroides strain {WS8N}.
\newblock Journal of Bacteriology 193: 4027--4028.
\bibAnnoteFile{porter}

\bibitem{pine}
 (Accessed 2013-04-01).
\newblock Pine reference sequences.
\newblock http://www.pinegenome.org/pinerefseq/.
\bibAnnoteFile{pine}

\bibitem{henson}
Henson J, Tischler G, Ning Z (2012) Next-generation sequencing and large genome
  assemblies.
\newblock Pharmacogenomics 13: 901--915.
\bibAnnoteFile{henson}

\bibitem{li}
Li H, Ruan J, Durbin R (2008) Mapping short {DNA} sequencing reads and calling
  variants using mapping quality scores.
\newblock Genome Research 18: 1851--1858.
\bibAnnoteFile{li}

\bibitem{trimble}
Trimble WL, Keegan KP, {D'Souza} M, Wilke A, Wilkening J, et~al. (2012)
  Short-read reading-frame predictors are not created equal: sequence error
  causes loss of signal.
\newblock {BMC} bioinformatics 13: 183.
\bibAnnoteFile{trimble}

\bibitem{schaper}
Schaper E, Kajava AV, Hauser A, Anisimova M (2012) Repeat or not
  repeat?--statistical validation of tandem repeat prediction in genomic
  sequences.
\newblock Nucleic Acids Research 40: 10005--10017.
\bibAnnoteFile{schaper}

\bibitem{zerbino}
Zerbino DR, {McEwen} GK, Margulies EH, Birney E (2009) Pebble and rock band:
  Heuristic resolution of repeats and scaffolding in the velvet short-read de
  novo assembler.
\newblock {PLoS} {ONE} 4: e8407.
\bibAnnoteFile{zerbino}

\bibitem{alba}
Alba E, Luque G (2007) A new local search algorithm for the {DNA} fragment
  assembly problem.
\newblock Evolutionary Computation in Combinatorial Optimization : 1–12.
\bibAnnoteFile{alba}

\bibitem{boetzer}
Boetzer M, Henkel CV, Jansen HJ, Butler D, Pirovano W (2010) Scaffolding
  pre-assembled contigs using {SSPACE}.
\newblock Bioinformatics 27: 578--579.
\bibAnnoteFile{boetzer}

\bibitem{dayarian}
Dayarian A, Michael TP, Sengupta AM (2010) {SOPRA:} scaffolding algorithm for
  paired reads via statistical optimization.
\newblock {BMC} Bioinformatics 11: 345.
\bibAnnoteFile{dayarian}

\bibitem{gritsenko}
Gritsenko AA, Nijkamp JF, Reinders MJT, Ridder Dd (2012) {GRASS:} a generic
  algorithm for scaffolding next-generation sequencing assemblies.
\newblock Bioinformatics 28: 1429--1437.
\bibAnnoteFile{gritsenko}

\bibitem{huson}
Huson DH, Reinert K, Myers EW (2002) The greedy path-merging algorithm for
  contig scaffolding.
\newblock Journal of the {ACM} ({JACM)} 49: 603–615.
\bibAnnoteFile{huson}

\bibitem{kececioglu}
Kececioglu JD, Myers EW (1995) Combinatorial algorithms for {DNA} sequence
  assembly.
\newblock Algorithmica 13: 7--51.
\bibAnnoteFile{kececioglu}

\bibitem{pop2004}
Pop M, Kosack DS, Salzberg SL (2004) Hierarchical scaffolding with bambus.
\newblock Genome Research 14: 149--159.
\bibAnnoteFile{pop2004}

\bibitem{delorme}
Delorme C, Poljak S (1993) Laplacian eigenvalues and the maximum cut problem.
\newblock Mathematical Programming 62: 557–574.
\bibAnnoteFile{delorme}

\bibitem{sokal}
Sokal R, Michener C (1958) A statistical method for evaluating systematic
  relationships.
\newblock Univ Kans Sci Bull 38: 1409--1438.
\bibAnnote{sokal}{The following values have no corresponding Zotero {field:ID}
  - Sokal-Michener-1958-{UKSBL3} - citeulike-article-id:7655015}

\bibitem{steinbach}
Steinbach M, Karypis G, Kumar V, et~al. (2000) A comparison of document
  clustering techniques.
\newblock In: {KDD} workshop on text mining. volume 400, p. 525–526.
\newblock
  \urlprefix\url{ftp://wilbur.eng.auburn.edu/pub/ezs0009/clustering/10.1.1.125.9225.pdf}.
\bibAnnoteFile{steinbach}

\bibitem{rizzo}
Szekely GJ, Rizzo ML (2005) Hierarchical clustering via joint between-within
  distances: Extending ward's minimum variance method.
\newblock Journal of Classification 22: 151--183.
\bibAnnoteFile{rizzo}

\bibitem{shen}
Shen H, Cheng X, Cai K, Hu MB (2009) Detect overlapping and hierarchical
  community structure in networks.
\newblock Physica A: Statistical Mechanics and its Applications 388:
  1706--1712.
\bibAnnoteFile{shen}

\bibitem{ward}
Ward JH (1963) Hierarchical grouping to optimize an objective function.
\newblock Journal of the American Statistical Association 58: 236--244.
\bibAnnote{ward}{The following values have no corresponding Zotero {field:PB} -
  American Statistical Association}

\bibitem{murtagh}
Murtagh F (1983) A survey of recent advances in hierarchical clustering
  algorithms.
\newblock The Computer Journal 26: 354 --359.
\bibAnnoteFile{murtagh}

\bibitem{ausiello}
Ausiello G, Firmani D, Laura L (2012) Real-time monitoring of undirected
  networks: Articulation points, bridges, and connected and biconnected
  components.
\newblock Netw 59: 275--288.
\bibAnnoteFile{ausiello}

\bibitem{hopcroft}
Hopcroft J, Tarjan R (1973) Algorithm 447: efficient algorithms for graph
  manipulation.
\newblock Commun {ACM} 16: 372--378.
\bibAnnoteFile{hopcroft}

\bibitem{khuller}
Khuller S, Raghavachari B (2010) Basic graph algorithms.
\newblock In: Atallah MJ, Blanton M, editors, Algorithms and theory of
  computation handbook, Chapman \& Hall CRC. pp. 7--7.
\bibAnnoteFile{khuller}

\end{thebibliography}

\end{document}